# Concentration of research funding leads to decreasing marginal returns

Philippe Mongeon[1], Christine Brodeur[2], Catherine Beaudry[3] and Vincent Larivière[1,4]

[1]École de bibliothéconomie et des sciences de l'information, Université de Montréal
C.P. 6128, Succ. Centre-Ville, Montréal, QC. H3C 3J7 Canada

[2]Bibliothèque, École Polytechnique de Montréal
C.P. 6079, Succ. Centre-Ville, Montréal, QC. H3C 3A7 Canada

[3]Département de mathématiques et de génie industriel, École Polytechnique de Montréal
C.P. 6079, Succ. Centre-Ville, Montréal, QC. H3C 3A7 Canada
Centre Interuniversitaire de Recherche sur la Science et la Technologie (CIRST)
Université du Québec à Montréal
CP 8888, Succ. Centre-Ville, H3C 3P8 Montreal, Qc. (Canada)

[4]Observatoire des Sciences et des Technologies (OST), Centre Interuniversitaire de Recherche sur la Science et la Technologie (CIRST), Université du Québec à Montréal
CP 8888, Succ. Centre-Ville, H3C 3P8 Montreal, Qc. (Canada)

**Corresponding author**

Philippe Mongeon
philippe.mongeon@umontreal.ca

**Acknowledgements**

This work is funded by the Canada Research Chairs program, as well as by the Social Sciences and Humanities Research Council of Canada. The authors also wish to thank the two anonymous referees for their useful comments and suggestions.

## Abstract

In most countries, basic research is supported by research councils that select, after peer review, the individuals or teams that are to receive funding. Unfortunately, the number of grants these research councils can allocate is not infinite and, in most cases, a minority of the researchers receive the majority of the funds. However, evidence as to whether this is an optimal way of distributing available funds is mixed. The purpose of this study is to measure the relation between the amount of funding provided to 12,720 researchers in Québec over a fifteen year period (1998-2012) and their scientific output and impact from 2000 to 2013. Our results show that both in terms of the quantity of papers produced and of their scientific impact, the concentration of research funding in the hands of a so-called 'elite' of researchers generally produces diminishing marginal returns. Also, we find that the most funded researchers do not stand out in terms of output and scientific impact.

## Introduction

In most countries, basic research is supported by governmental research councils that select, after peer review, the individuals or teams that are to receive funding. In Québec, for example, 70.5% of external research funding comes from three provincial and three national research councils (Robitaille & Laframboise 2013). Unfortunately, the budgets of these research councils are limited and, thus, a large proportion of grant proposals—as well as scholars—remain unfunded. For example, a previous study showed that 20% to 45% of Québec's researchers, depending on the discipline, had no external funding between 1999 and 2006 (Larivière et al. 2010). Along these lines, national scientific agencies, including the National Science Foundation (NSF – United States) and the Canadian Institutes of Health (CIHR) tend to give fewer grants of a higher value, which leads to high rejection rates (Joós 2012; CIHR 2012; NSF 2013) as well as to an increased concentration of available funds in the hands of a few researchers. More specifically, 10% of the researchers funded by the Social Sciences and Humanities Research Council of Canada (SSHRC) accumulate 80% of available funds, 10% of those funded by the Canadian Institutes of Health Research (CIHR) obtain 50% of the funds, and 10% of those funded by the NSERC accumulate 57% of the funds[1]. The situation is similar in Québec when we combine funding from the national and provincial agencies: 20% of researchers obtain 80% of the funds in social sciences and humanities (SSH), 50% of the funds in health, and 57% of the funds in natural sciences and engineering (NSE) (Larivière et al. 2010).

In an attempt to ensure that the funds are effectively used, the NIH started doing special analyses of applicants who are already receiving more than one million dollars in grants. At the National Institute of General Medical Sciences (NIGMS), this practice of special review for researchers who are already holding a large sum of funding has been going on for more than 20 years (Berg 2012). Still, with a few researchers receiving most of the funds available and many not receiving any, it seems legitimate to ask whether this concentration of funds leads to better collective gains than funding policies that promote a more even distribution of funding.

---

[1] Data compiled by the Observatoire des Sciences et Technologies (OST) using results of competition for each of the councils, and the *Almanac of Post-Secondary Education in Canada,* of the Canadian Association of University Teachers.

This study aims to contribute to the discussion on the optimal level of research funding by analysing the relationship between all funding received by all of Québec's researchers over a period of 15 years (1998-2012) and their research output and impact from 2000 to 2013. More specifically, it aims at answering two questions: 1) how do research productivity and scientific impact of individual researchers vary with the amount of funding they receive? 2) is this variation similar in the three general fields of science that are health, natural sciences and engineering, and social sciences?

## Theoretical framework

The scientific community is a field (i.e. a social space) involving agents that share a common goal (the production of knowledge) and compete against each other for their peers' recognition (Bourdieu 1976). In order to work efficiently (i.e., to maximise knowledge production), the scientific community developed a reward system which provides different forms of recognition to scientists who best achieved this goal (Merton 1957). This results in a social stratification of the scientific community (Cole & Cole 1973), which in turns leads to what Zuckerman (1998) described as an effect of accumulation of advantages and disadvantages. This process starts at the very beginning of scientists' careers, as those with the most potential are given more resources (e.g., obtain scholarships, get accepted in prestigious departments, etc.) which will have both short and long term effects on their scientific career and thus will help acquire even more resources. When such a system is working optimally, it should yield exponential returns in terms of knowledge production since the best researchers have more resources and can, moreover, use them more efficiently than their colleagues with fewer resources (Zuckerman 1998). If we consider research funding, the reward system should, in theory, give more resources to the best researchers who will, in turn, produce more or better output.

However, some elements within of the reward system itself may have a negative impact on its efficiency, one of them being the Matthew effect described by Merton (1968). As highlighted by Laudel (2005), funding decisions are based not only on the quality of the research proposal but also on the quality of the researchers, which is to a large extent assessed by the researcher's past accomplishments. We may thus wonder if, and to what extent, research funding becomes less of an instrumental reward, geared towards maximizing knowledge production, and more of an honorific reward. According to Schmoch & Schubert (2009), this shift is a consequence of the growing scarcity of research funds.

Also, in practice, a number of factors (e.g., limited time, administrative and teaching responsibility) may limit individual researchers' productivity, so that such exponential returns can hardly be expected. Larger projects requiring more resources and more people may also require more coordination effort, and getting more funding also requires spending more time writing grant proposals. Thus, we might expect that lower amounts of funding will yield greater returns in terms of scientific output and impact, and that these returns will start decreasing with higher amounts of funding.

## Literature review

Despite the central role of research funding in the current research system, its relation to research outcomes has not been extensively studied. Some studies have looked at the influence of funding—using funding acknowledgements as an indicator—on the scientific impact of articles. For instance, Zhao (2010) analysed a small sample of articles published in 7 core LIS journal and found that funded articles had, in general, a higher impact and, on average, were cited faster than non-funded articles. She, however, observed that the two most cited articles did not disclose any funding. A similar study using a subset of Iranian publication concluded that the average citation rate of funded articles (as determined by acknowledgements) is higher than that of unfunded papers (Jowkar et al. 2011). Other studies have examined the impact and productivity of scientists who see their application granted versus the ones whose applications are rejected. Two studies, one from Canada (Campbell et al. 2010) and the other from the Netherlands (van Leeuwen & Moed 2012), found that researchers who receive a grant tend to have higher citation impact than applicants who did not receive funding. A third study, focused more on research productivity than scientific impact, found that professors with external funding published more than the ones with no external funding (Gulbrandsen and Smeby 2005). These studies suggest that receiving funding has a positive effect on a researcher's scientific output.

Other studies have investigated the relationship between the amount of funding received and research productivity and scientific impact, first at high levels of aggregation. Towards the end of the 1970's, the relationship between R&D expenditure and the number of papers was investigated at the level of universities, for 11 different fields (McAllister and Wagner 1981). A strong indication of a linear relationship was found. In the 1990's, Moed and colleagues (1998) analysed the productivity of Flemish university departments, finding that those with the highest externally funded capacity[2] had a strong decrease in publication productivity. More recently, the effect of the amount of funding was studied at the individual level. Heale et al. (2004) reported that one of the strongest determinants of the number of papers published was the amount of funding received. Moreover, they concluded that an increase in funding was associated with an increase in the number of articles, but not proportionally. Nag et al. (2013) reached a similar conclusion regarding the number of publications after studying a sample of 720 American bioscientists. More precisely, the authors calculated that increasing a laboratory's budget by 10% lifts article output by only 7.5%, thus suggesting decreasing marginal returns. Fortin and Currie (2013) also found decreasing marginal returns when they looked at the number of publications of Canadian researchers in Integrative Animal Biology, Inorganic & Organic Chemistry and Evolution & Ecology who received NSERC funding. Along these lines, Berg (2010) observed that the number of publications of NIGMS grantees did not increase linearly with the amount of funding they received, but rather appeared to reach a plateau (Berg 2010). Lastly, a few studies discussing the validity of external funding as an indicator to evaluate researchers have also found that the output typically rises with funding until a point where it reaches a plateau and eventually starts decreasing (e.g., Laudel 2005; Schmoch & Schubert 2009; Schmoch et al. 2010). For low values of third-party funds, increasing them also increases efficiency. However, when the funding reaches a certain threshold value, its efficiency stops increasing and starts decreasing (Schmoch & Schubert 2009). On the whole, evidence on the relationship between research funding and various output of research does not converge—mainly because of different methods used and population studied—although many papers seem to suggest that the relationship between funding and research outputs is not linear.

---

[2] Moed et al (1998) define the research capacity as the number of full-time equivalents (FTE) spent on scientific research, the *externally funded* research capacity being funded by external sources (i.e. other than the researchers' basic allowance).

# Data

Data on funding for all Québec's academic researchers from 1998 to 2012 were obtained from the Information System on University Research (SIRU), an administrative database from the Québec provincial government that covers all funded research in Québec's universities. Access to this data was provided by the *Observatoire des sciences et des technologies* (OST). The funding data includes different types of funding (e.g., grants, contracts and internal funding), but we chose to include only grants as they represent the bulk of university funding and are awarded through peer-review. Those grants were given out by more than 900 hundred organizations and amounted to $8.1 Billion, most funds (about 50%) being given out by the three Federal research councils of Canada (the Social Sciences and Humanities Research Council [SSHRC], the Natural Sciences and Engineering Research Council [NSERC] and the Canadian Institutes of Health Research [CIHR]), as well as the three research councils in Québec (The Fonds de recherche du Québec – Société et culture [FRQ-SC], the Fonds de recherche du Québec – Nature et Technologie [FRQ-NT] and the Fonds de recherche du Québec – Santé [FRQ-S]).

Publication data for each researcher were obtained from Thomson Reuters' Web of Science for the 2000-2013 period. The Web of Science data was also provided by the OST, which maintains up to date disambiguated publication data for all Québec researchers since 2000. Then, researchers were divided into four broad research disciplines: Arts and Humanities (AH), Social Sciences (SS), Natural Sciences and Engineering (NSE) and Health. Researchers were associated with the discipline in which most of their papers are published. Researchers for which no publications were found as well as those with the same number of publications in two or more disciplines were categorized according to their department. We decided to exclude AH researchers from our analysis considering the poor coverage of Arts and Humanities publications in the Web of Science described in previous studies (e.g., Larivière et al. 2006; Mongeon & Paul-Hus 2015). Therefore, this paper focuses on the remaining three disciplines: SS, NSE and Health. The number of researchers in each field, as well as the total number of papers and citations, is shown in Table 1. For each researcher, we calculated the total amount of funding received. The total funds attributed to each project were divided by the number of researchers on the application, each of them receiving an equal share.

Three indicators were used to measure research outcomes of researchers: the total number of articles, the average relative citations (ARC) and the number of top cited articles (top 10% most cited in the sub-discipline). The ARC of individual papers is calculated by dividing the number of citations received by the paper by the average number of citations received by all papers of the same sub-discipline published in the same year. The sub-discipline is defined by the journal in which the paper is published, and the journal classification is based on the NSF classification which contains 144 sub-disciplines. This allows us to account for the time of publication and disciplinary differences in terms of citations. The amount of funding received as well the five outcome indicators were annualized to be able to compare researchers who were not active at the same time over the 1998 and 2013 period. To do this, we looked at the year of researchers' first grant and the year of their first publication. The earliest of these two years was considered the first active year of the researcher. For example, if a researcher received his first grant in 2002 and published an article in 1999, he or she was considered active since 1999. Similarly, researchers were considered active until the year of their last publication or grant. The number of researchers, papers and citations included in our dataset is shown in Table 1.

**Table 1. Funded researchers, papers published and citations received in Québec**

| Discipline | Number of researchers | Funded researchers N | % | Articles | Citations | Top cited articles |
|---|---|---|---|---|---|---|
| Health | 4,742 | 3,477 | 73.3% | 65,131 | 1,992,894 | 10,063 |
| NSE | 3,142 | 2,763 | 87.9% | 51,792 | 885,680 | 7,306 |
| SS | 4,836 | 3,877 | 80.2% | 13,725 | 184,826 | 1,608 |
| Total | 12,720 | 10,117 | 79.4% | 130,648 | 3,063,400 | 18,977 |

## Analysis

**Marginal returns**
We used the Cobb-Douglas production function to see whether the annual research funding of researchers yields increasing, constant or decreasing marginal returns. Since funding is the only independent variable in our model, the Cobb-Douglas function takes the form $ln(Q) = ln(\alpha) + \beta(ln(K))$ where $Q$ is the output in terms of articles, ARC or Top cited articles, and $K$ is the input in research funding. The intercept ($\alpha$) and the coefficient ($\beta$) are calculated by running a linear regression on the natural logarithm of funding and the measured indicator. A $\beta$ lower than 1 indicates decreasing marginal returns, a $\beta$ equal to 1 indicates constant marginal returns and a $\beta$ higher than 1 indicates increasing marginal returns.

We used local regression (LOESS) to visualize the increasing, constant or diminishing marginal returns of research funding. The LOESS function fits a polynomial for each point in a scatterplot using weighted least squares on a subset of the data. More weight is given to the closest neighbours and less to the data points that are further away. We used a 50% smoothing parameter (50% of the data is used to fit the polynomial on each data point), to obtain a smooth curve that is not overly affected by normal variation in the data.

**Outlier analysis**
A first visual of our data displayed some obvious outliers (e.g., heavily funded under-performers and minimally funded over-performers). To ensure that our analyses using the Cobb-Douglas and LOESS functions were robust, we performed the analyses on the complete dataset as well as in the dataset with the outliers removed. Outliers were identified statistically rather than manually, by running a linear regression on the data and calculating the Cook's distance for each data point. We identified as outliers the points with a Cook's distance higher than a limit obtained with formula 4/($n$-$k$-1), where $n$ is the number of data points and $k$ is the number of independent variables.

## Results

**Distribution of research funding**
As Figure 1 shows, research funding provided to Québec researchers over the 1998-2012 period is concentrated in the hands of a minority of researchers. In SS and health, about 20% of the researchers shared 80% of all funding given during that period. The funds were less concentrated in NSE, where 25% of the researchers shared 80% of the funds, than in the other fields. On the whole, these results confirm that research funds in Québec are indeed concentrated in the hands of a few researchers.

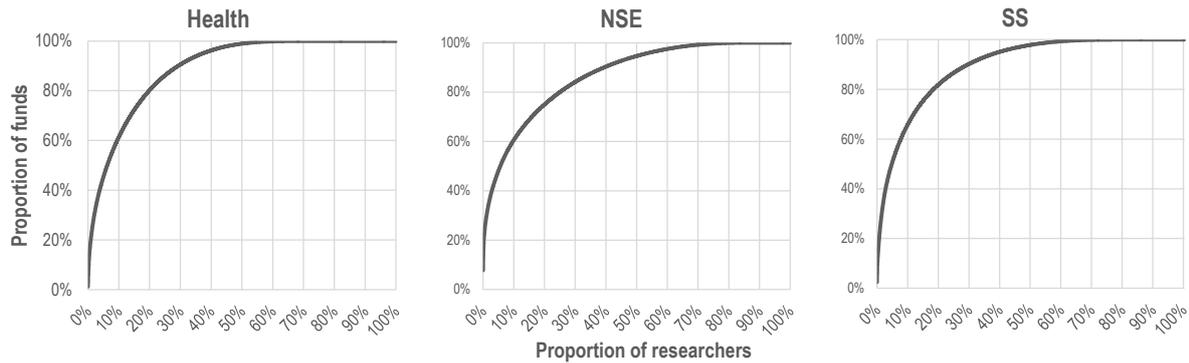
**Figure 1. Distribution of research funds among all researchers (1998-2012)**

Since we used articles published before 2013, our results could be affected if a disproportionately large amounts of funding had been awarded in recent years when the full output from those grants (in terms of both productivity and impact) will not yet have been achieved. Thus, we looked the evolution of the total amount of grants received by Québec's researchers over the 1998-2012 period (Figure 2). We observe that funding in Health as greatly increased between 1998 and 2005, and has been relatively stable since then. In NSE, we observe a steady increase, going from 81 to 239 million $CAN over the same period. SS researchers received less funding than their colleagues in Health and NSE. Their total funding increased from 33 to 72 million $CAN between 1998 and 2004 and has been stable since then.

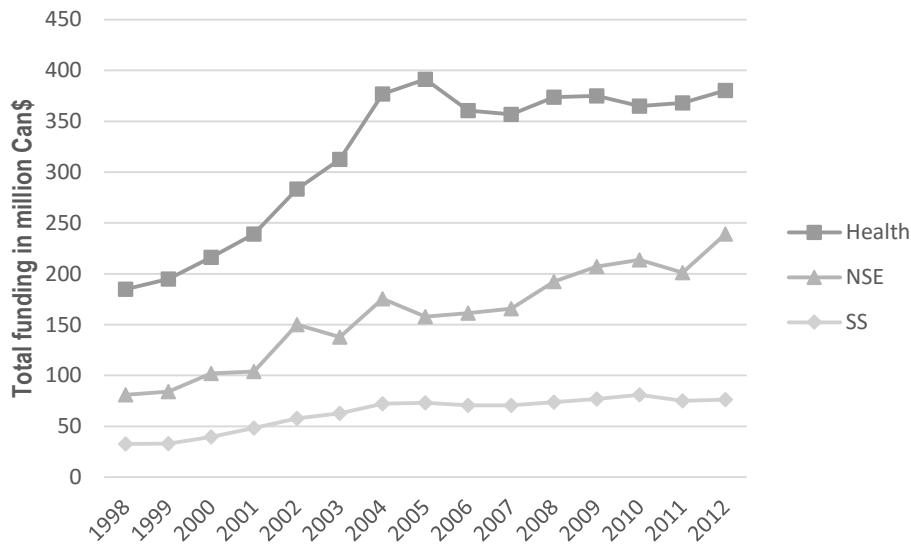
**Figure 2. Total amount of funding by discipline (1998-2012)**

We also looked at the 50 largest grants in each field to see if a larger share of them were obtained in the 2010-2013 period. Since it was not the case, it seems reasonable to conclude that the possibly missing output from grants awarded in the last years of the period will not affect our results.

**Marginal returns of research funding**
The results of the Cobb-Douglas function used to determine whether increased research funding yields increasing, constant or decreasing marginal returns, are shown in Table 2. It provides the results in terms of the number of articles, the ARC and the number of top cited articles for each field, as well as for both the complete data set and the data set without outliers.

**Table 2. Relation between the amount of funding and the research output and impact as measured by the Cobb-Douglas production function.**

|  |  |  |  | Complete data set | | | | | Without outliers | | | |
|---|---|---|---|---|---|---|---|---|---|---|---|---|
|  |  | N |  | Coefficient | Standard Error | t Stat | P-Value | N | Coefficient | Standard Error | t Stat | P-Value |
| Health | Articles | 4742 | Intercept | -2.723 | 0.052 | -52.451 | 0.000 | 4570 | -4.355 | 0.216 | -20.186 | 0.000 |
|  |  |  | Funding | 0.252 | 0.005 | 48.845 | 0.000 |  | 0.404 | 0.021 | 19.144 | 0.000 |
|  | ARC | 4742 | Intercept | -3.053 | 0.052 | -58.187 | 0.000 | 4657 | -3.493 | 0.222 | -15.709 | 0.000 |
|  |  |  | Funding | 0.233 | 0.005 | 44.602 | 0.000 |  | 0.273 | 0.022 | 12.676 | 0.000 |
|  | Top cited articles | 4742 | Intercept | -6.065 | 0.057 | -106.299 | 0.000 | 4579 | -10.351 | 0.317 | -32.608 | 0.000 |
|  |  |  | Funding | 0.240 | 0.006 | 42.370 | 0.000 |  | 0.643 | 0.031 | 20.705 | 0.000 |
| NSE | Articles | 3142 | Intercept | -4.049 | 0.087 | -46.566 | 0.000 | 3082 | -10.214 | 0.402 | -25.432 | 0.000 |
|  |  |  | Funding | 0.321 | 0.009 | 37.112 | 0.000 |  | 0.924 | 0.040 | 23.289 | 0.000 |
|  | ARC | 3142 | Intercept | -4.691 | 0.095 | -49.498 | 0.000 | 3066 | -10.567 | 0.458 | -23.076 | 0.000 |
|  |  |  | Funding | 0.295 | 0.009 | 31.359 | 0.000 |  | 0.868 | 0.045 | 19.230 | 0.000 |
|  | Top cited articles | 3142 | Intercept | -7.335 | 0.103 | -71.108 | 0.000 | 3082 | -17.351 | 0.509 | -34.059 | 0.000 |
|  |  |  | Funding | 0.242 | 0.010 | 23.569 | 0.000 |  | 1.220 | 0.050 | 24.239 | 0.000 |
| SS | Articles | 4836 | Intercept | -6.123 | 0.071 | -86.189 | 0.000 | 4705 | -10.425 | 0.398 | -26.222 | 0.000 |
|  |  |  | Funding | 0.192 | 0.008 | 24.999 | 0.000 |  | 0.644 | 0.043 | 14.842 | 0.000 |
|  | ARC | 4836 | Intercept | -6.554 | 0.073 | -89.714 | 0.000 | 4723 | -11.496 | 0.412 | -27.935 | 0.000 |
|  |  |  | Funding | 0.197 | 0.008 | 24.856 | 0.000 |  | 0.718 | 0.045 | 16.077 | 0.000 |
|  | Top cited articles | 4836 | Intercept | -8.521 | 0.047 | -182.785 | 0.000 | 4706 | -11.521 | 0.252 | -45.729 | 0.000 |
|  |  |  | Funding | 0.079 | 0.005 | 15.599 | 0.000 |  | 0.380 | 0.027 | 13.844 | 0.000 |

For the complete dataset, the coefficients are all below 1, which means that, in each field and for each indicator, research funding yields decreasing marginal returns. Furthermore, this decrease is quite sharp given that the values of the coefficients are very small, ranging from 0.079 for the top 10% most cited articles in social sciences to 0.321 for the number of articles in Natural Sciences and Engineering. However, outliers appear to have and important effect on these coefficients, especially in NSE and Health. Indeed, while we can still observe coefficients below 1 for most indicators once we remove the outliers, the coefficient are much higher for those fields, and it even becomes positive for top cited articles in NSE. Thus, in this field, the decreasing marginal returns observed in terms of top cited articles might be caused by the presence of highly influential outliers, who obtain a lot of funding and publish an important number of important papers.

Figure 3 shows the scatterplots with a fitted a LOESS curve for each field and indicator. It excludes outliers, as they affect the trends. However, the outlier themselves might provide interesting insights on the relation between funding and performance, which will be discussed later in the paper. The distribution of the points on the graphs show a few interesting elements. First, the points are very densely concentrated on the left of the y-axis, and the density decreases as the funding increases. This is coherent with the funding landscape presented above, with most researchers obtaining relatively small amounts of funding, and around 20% obtaining no funding at all. Another interesting element is that researchers who get the highest scores for each indicator are also situated on the left of the y-axis. Conversely, apart from a few exception, researchers who receive large amounts of funding do not appear to get particularly high scores. Also, the scatterplots show a high variability in the indicators at any level of funding, which suggest that the performance of researchers depends only partially on the funding they receive and that other factors are likely to play an important role.

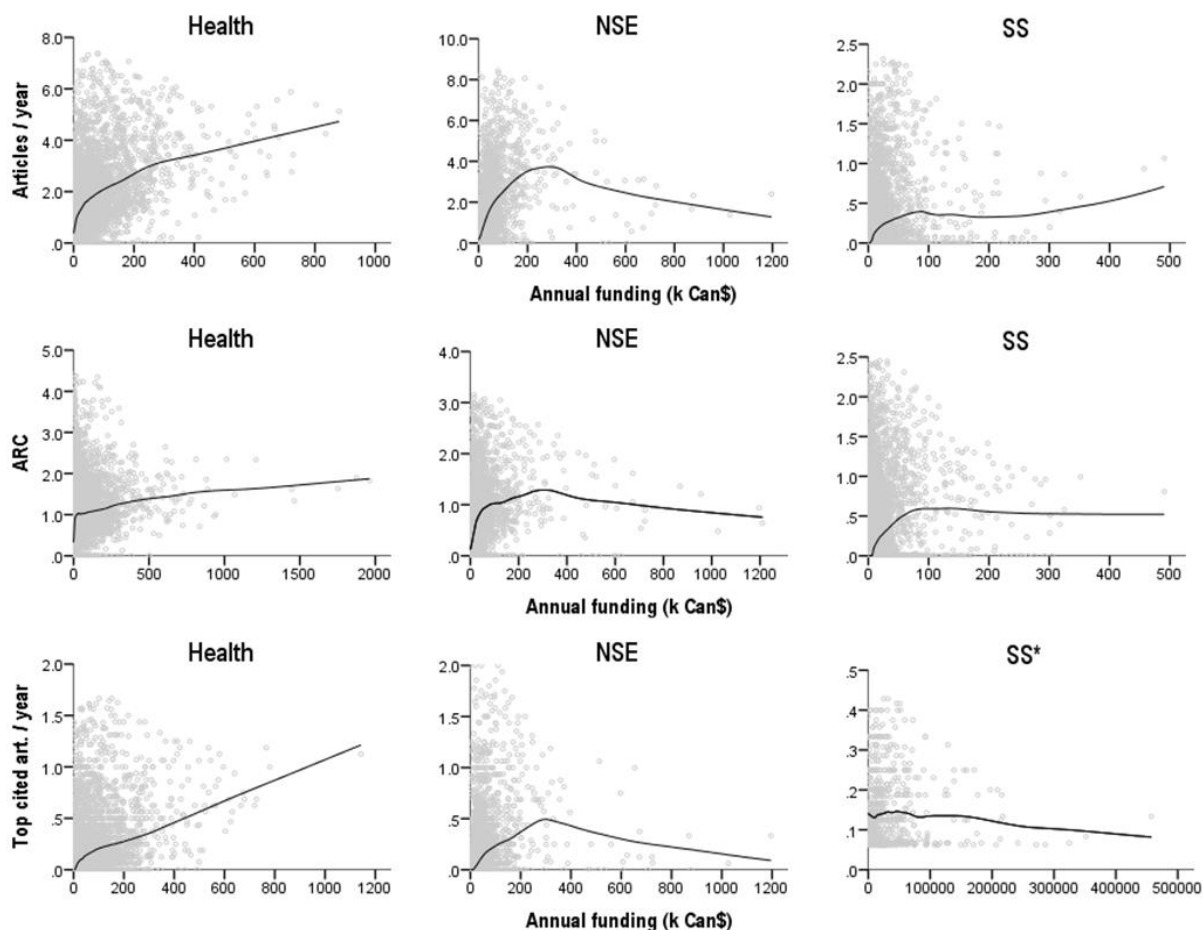

**Figure 3. Relation between the amount of funding and the research output and impact.**
* The vast majority of researchers in SS have no top cited articles. To avoid the LOESS curve to be pulled to 0, we kept only researchers with at least 1 top cited article.

The LOESS curves display decreasing marginal returns similar to what was found using the Cobb-Douglas production function. In almost all cases, the curves show an important increase of the measured indicator at the lowest amount of funding, followed by a slower increase once a specific threshold is reached. In NSE, the curve even shows a decline at higher amounts of funding. One exception is the top cited articles indicator for researchers in SS, which shows a rather constant decline as funding increases. This suggests that funding might, in fact, have little effect on the production of top cited articles in Social Sciences. Health also stands out from the other two fields, as the LOESS curve shows an increase throughout the graph. Thus, if we take the disciplines globally, these results suggest that decreasing marginal returns could be more important in NSE with an eventual decline of the indicators. The decreasing returns would be less important in Health, where the indicators keep rising while SS are in between with a quick rise followed by a more or less horizontal LOESS curve.

While it is not the purpose of this paper to determine an optimal amount of funding, the LOESS curves do provide some indications on the amount of funding that yields the highest return per dollar in each field. For example, in SS and Health, the first break in the line appears at around 50,000 $CAN. Here, the ARC in Health is an exception: it seems that the fact of being funded or not makes an important difference while the amount of funding has a much lower effect on the indicator. In NSE, the LOESS curve shows two breaks. The first one, similarly to Health and SS, occurs at around 50,000 $CAN annually. The curves then keep rising until they reach a point where they start decreasing, at around 300,000 $CAN annually. Some potential explanations for this specificity of NSE are provided in the discussion below.

## Discussion

Our results show that funding is strongly related with the research productivity and scientific impact of individual researchers, although there are decreasing marginal returns for most of the indicators. Indeed, the number of articles, the average relative citations and the number of top cited articles showed a non-proportional increase with an increase in funding, until a certain point where they start to decrease. Moreover, when both inputs and outputs increase, the growth of outputs is not proportional to that of inputs, and decreasing marginal returns can be observed in that part of the LOESS curve. For the average relative citation rates (ARC), we observe a rapid growth until a plateau is reached. This also implies decreasing marginal returns.

Most of the results presented above did not include outliers because of their potentially high influence on the trends observed. However, outliers provide important insights on the relationship between funding and outcomes. For example, many of the outliers were the most funded researchers in their field. More specifically, the 20 most funded researchers (who shared 13% of the total funding in their field over the whole 15 years period) in Health were kept out of the analysis, same for the 20 most funded scholars in NSE and SS, who shared 22% and 14% of the total funding in the field, respectively. On average these outliers received 27 times, 40 times and 32 times more funding than the rest of the researchers in Health, NSE and SS respectively, yet they "only" published on average 6, 5 and 19 times more than their average colleague. Moreover, from a scientific impact point of view, the 20 most funded scholars in Health and SS obtained citation rates that are respectively 1.68 and 1.15 times greater than the average scholar of their discipline, while they had on average a slightly lower ARC (0.732) than their average colleague (0.786) in NSE. On the other hand, many researchers with low funding but high output or impact were also identified as outliers. These two groups of outliers help to highlight the fact that having a lot of funding does not necessarily lead to a higher output or impact and that even researchers with low amounts of funding can achieve a high level of output and impact. Hence, at least from a bibliometric standpoint, this concentration of research funding in the hands of a minority of researchers brings no clear collective advantages in terms of output and impact, suggesting that such funding policies might not be efficient. Indeed, as decreasing marginal returns are observed, one may wonder what could justify awarding millions of dollars to a few researchers while many others receive nothing. Thus, in terms of funding policy, our results, along with those of Fortin and Currie (2013), support an approach where a higher number of smaller grants are given to a higher number of researchers rather than one where the funds are concentrated among a minority of researchers who obtain very large grants.

There are many factors that could explain some of the decreasing marginal returns observed. One of them might be a change in the amount of time researchers have to spend on tasks other than research. For instance, one might argue that drafting grant proposals takes time and that obtaining higher amounts of funding might require researchers to spend more time writing grant proposals and less time performing research. Furthermore, highly funded researchers might be leading larger projects that require more coordination effort and time, which also reduce the amount of time left to do research. Moed et al. (1998) observed, in the 1980s, a decrease in the number of publications from Flemish universities that had the strongest increase in competitively acquired funding. They suggest that this might be because highly funded departments attract a large number of young scientists, which may require more supervision from the departments' senior scientists. This reduces the time such experienced researchers (i.e. potentially more productive) spend on research, therefore decreasing their output. Similarly, Moed et al. (1998), pointed out that young foreign scientists were more likely to work in the highly funded departments. They suggested that those scientists were typically visiting for short periods and mostly for learning rather than publishing articles, which might, in turn, result in lower time for research for senior researchers because of their supervisory role. Some research projects may also require important investments in equipment or infrastructure, which do not necessarily lead to an increased output or impact. However, this factor should have little effect on the results of this study because, in Canada, infrastructure investments are typically funded by a specific organisation (Canadian Foundation for Innovation), and those grants were not included in our data. Furthermore, as Laudel (2005) suggests, the quality of a research proposal and the past performance of a researcher do not guarantee the success of external funding, which is also determined by factors that have little to do with quality and performance. Therefore, while receiving funding does provide researchers with the means to carry on their research projects, it does not guarantee that they will succeed at achieving publishable results. Also, a research project may simply be failing despite researchers asking for more and more funds to make it succeed.

Another possibility is that, as we hypothesised, there is a Matthew effect in research funding. In a context where competition for research funding and the concentration of funds are both increasing, researchers who have acquired the esteem of their community from their past achievements have more facility than others to obtain funding, and moreover, to obtain larger ones. However, a large number of previous studies have found, researchers tend to be more productive early in their career (e.g., Cole 1979; Dennis 1956; Lehman 1953; Levin & Stephan 1991). This could partly explain the decreasing marginal returns observed. Also, research grants are sometimes used as a performance indicator, which encourages researchers to apply for more (and bigger) grants (Hornbostel 2001) that they might not necessarily need to perform their research. This could lead to an inefficient use of the funds received (Sousa 2008). Finally, we may be observing a shift from research funding as an instrumental reward to research funding as an honorary reward as we move towards higher levels of funding. Smaller amounts of funding would be given to promising researchers and projects that produce higher returns while receiving funding beyond this point would be more likely related to past performance. Furthermore, some highly funded researchers might have relatively low scores in terms of output and impact because their work has different characteristics and because they play a different role (one that cannot be observed using bibliometric methods) in the scientific community. For example, Schmoch & Schubert (2009) and Schmoch et al. (2010) divided researchers in four main groups (networkers, educators, high producers and high impact researchers), all of them being, in their view, essential to the sustainability of the scientific system. While high producers publish many articles and high impact researchers are focused on high impact articles, networkers and educators focus on different aspect of the academic life. The existence of these different types of researchers could explain some of the variability in the results in general, and it may also be the case that some researchers of the non-publication focused types are overrepresented in among the most funded researchers. It could also be the case that researchers receiving larger grants do not always participate directly in the research funded by those grants (Boyack & Jordan 2011). To assess the extent to which these factors could have affected our results, we searched the top 20 most funded researchers in our dataset on the Web and looked at their academic profile. A majority of them were older researchers with important administrative functions. This was especially the case in NSE, where many deans and heads of departments were found among the top funded researchers. One possibility that would require further investigation is that, for example, these researchers might participate in the grant applications, using their reputation to improve the applications' chances of success, while it is more their co-applicants who will use the funds to produce scientific output.

## Limitations

In the present study, we used the entire population of Québec university researchers to analyse the relation between funding and the scientific output and impact. Our results show that researchers with a moderate amount of funding seem to provide better returns in terms of output or impact per dollar. In the discussion, many hypotheses regarding the different factors that could explain these observations have been suggested. This is the main limitation of this study: it does not take into account some of those potentially important confounding factors. Thus, further research should validate some of the hypotheses raised in our discussion by taking into account other independent variables such as the academic age of researchers, their role in the scientific community, the size and composition of teams, and some specificities of the funded projects. This will require a different design based on a subsample of our population for which this additional data will be needed. Further qualitative research could also focus on highly funded researchers to better understand the way they work and participate in the scholarly communication process.

Other limitations of this study should also be acknowledged. Firstly, we only took into account the date at which funding was received and papers published to determine the period during which researchers were active and to annualize the indicators. We can conclude that funding is related with output and impact and that each indicator has in most cases a rapid growth followed by a slower growth and sometimes even a decline when funding increases. However, we cannot infer any causal mechanisms. Along these lines, some of the potential outcomes of funding and research cannot be measured with bibliometric indicators (e.g., the number of students trained and social outcomes). Also, the funding received is sometimes linked to a particular project, and our study does not focus on the research output linked such specific grants. Thus, further research could aim at comparing research output and impact of funded projects specifically using funding acknowledgements, for example. Our study also didn't look at indicators of technology transfer (i.e. patents). As Bolli and Somogyi (2011) have found, technology transfer activities can have an impact on the relation observed between third party funding and basic research output. Another limitation comes from the lower coverage of the SS research output in the Web of Science (Archambault et al. 2006; Larivière et al. 2006; Mongeon & Paul-Hus 2015). Since researchers in SS are more likely than those in Health or NSE to write books or publish in local journals not included in the Web of Science, a significant part of their output could not be taken into account in this study. Furthermore, the timeframe of the analysis (1998-2013) involves that not all articles had the same time to accumulate citations. An article published in 2001, for example, might have accumulated more citation than a more recent article just because it had more time to do so.

## Conclusion

In a context where financial resources devoted to research are declining in constant dollars in many countries, it is important to ask whether the manner through which research funding is allocated is optimal. The results presented in this paper suggest that it might not be the case. Both in terms of the quantity of papers produced and of their scientific impact, the concentration of research funding in the hands of a so-called 'elite' of researchers produces diminishing marginal returns. From a policy perspective, this suggests that even though more funding does in general lead to a higher number of publications, giving bigger grants to fewer individuals may not be optimal. If the objective is to maximize output, then giving smaller grants to more researchers seems to be a better policy. In terms of scientific impact, the quickly reached plateau indicates that increasing funding has a very small effect on the average relative citations. Again, if the goal of research funding is to generate research that has a greater impact overall, our results suggest that giving smaller grants to a larger number of researchers may be a better decision, leading to more publications and citations overall. Thus, our results support the idea that a more egalitarian distribution of funds would yield greater collective gains for the scientific community. However, it should be stressed here that research output and impact do not necessarily equate with societal impact, so our findings do not support the claim that a more egalitarian distribution of funds would be better for society as a whole. Nonetheless, it should be understood that the main determinant of scientific production is not so much the money invested but rather the number of researchers at work (Abt 2007), and that by funding a greater number of researchers, we increase the overall research productivity. Furthermore, there is a certain degree of serendipity associated with scientific discoveries and funding the work of as many researchers as possible increases the likelihood that some of them make major discoveries.